\documentstyle[eqsecnum,aps,multicol,graphicx]{revtex}
\begin{document}
\draft
\title{Local properties and Density of States in the two-dimensional p-d Model}
\author{V. Fiorentino, F. Mancini, and E. \v Z\c asinas\cite{byline}}
\address{Dipartimento di Scienze Fisiche ``E.R. Caianiello'' -
Unit\`{a} INFM di Salerno, \\ Universit\`{a} degli Studi di
Salerno, 84081 Baronissi (SA), Italy }
\author{A. F. Barabanov}
\address{
Institute for High Pressure Physics, 142098 Troitsk, Moscow
Region, Russia}
\date{\today}
\maketitle
\begin{abstract}
The p-d model which well describes the CuO$_{2}$ planes of the
high-$T_c$ superconductors is studied by means of the Composite
Operator Method (COM). The relevant quasi-particle excitations are
represented by composite operators. As a result of taking into
account spin excitations we find a $p-$ like band near the Fermi
level. The dispersion of this band gives a Fermi surface which is
in good agreement with the experimental measurements. Due to the
strong mixing of the relevant excitations, the spectral weight of
this band is reduced and gives a large Fermi surface in the
moderately doped region. The dependence of the calculated physical
quantities on model parameters, temperature and doping, is in a
very good agreement with the available Quantum Monte Carlo
results.
\end{abstract}


\begin{multicols}{2}
\narrowtext

\section{Introduction}

Many unexplained exotic features of transition metal oxides
containing CuO$_{2}$ planes \cite{review1,review2} have inspired
the developing of various theoretical methods to solve, at least
approximately, the models which describe such compounds.

In the last years, a big effort has been devoted to explain
experimental results in terms of Hubbard-like or t-J models and
many interesting results have been successfully achieved. The
Hubbard model was suggested \cite {anderson,zhangrice} as low
energy derivative of the Emery (or p-d) three band model \cite
{emery}. The Emery model takes into account the crystalline
structure of CuO$_{2}$ planes and initially retains the orbitals
of $p$ and $d$ electrons at oxygen and copper sites, the $d$
orbital giving the strong Coulomb repulsion when occupied by two
electrons. Due to the more detailed structure of the p-d model,
one can expect that its solutions can give us a more rich physical
picture of cuprates and can explain a wider range of experimental
results than solutions of Hubbard-like or t-J models.

It is believed that the strong correlation of electrons in the copper $d$
orbitals leads both to the insulating antiferromagnetic state at low density
of doping carriers and to superconductivity at higher doping. The doping
dependence of the electronic state is not rigid band-like. This suggests
that the doped hole changes the state of nearby localized copper $d$
electron and reduces the energy of the system if compared to the case of
rigid band doping.

A detailed mechanism of binding $d$ electron to the doped hole was
proposed by Zhang and Rice \cite{zhangrice}. As a result of
binding, a local singlet forms and it can move through the
lattice. Further theoretical investigations of the band dispersion
of such singlet in CuO$_{2}$ planes gave a successful description
\cite{barabanov1,barabanov2} of ARPES results
\cite{tobin,abrikos,gofron,wels}.

However, to recover the doping and temperature dependence of micro- and
macro-scopic observables, one needs to study not only such singlet
excitations, although their energy lies near the Fermi level, but also the
wider basis of excitations which are possible in the plane.

Recently, a technique called the Composite Operator Method (COM)
has been proposed \cite{mancini2,mancini1,mancini5} to describe
the local and itinerant properties of strongly correlated systems.
A physical model is usually described in terms of elementary
particles and some interaction. However, at level of observation
the identity of the original particles is lost; the macroscopic
behavior of the system is described in terms of new excitation
modes. When the interaction is strong the properties of these
excitations will be very different from those of the original
particles and hardly obtainable by a perturbation scheme. By
following this scheme, in the COM a set of field operators is
taken as the basis in which the theory is developed. These fields
are chosen in order to describe the experimentally observed
properties and are called composite fields since they are
constructed from the initial set of particles. The properties of
these composite fields are dynamically determined by the
interaction and by the boundary conditions and must be self
consistently calculated. In this process a special attention is
put to preserve the symmetry properties of the model
\cite{mancini6,Adolfo}. Particularly, the conservation of the
Pauli principles plays a fundamental role. As discussed in Ref.\
\cite {mancini6}, although the method has many points in common
with the projection method \cite{becker}, the method of equation
of motion \cite{rowe} and the spectral density approach
\cite{kalashnikov}, remarkable differences arise according to the
different schemes of self consistency. The method has been applied
to the study of several models, like Hubbard \cite
{mancini1,mancini5,mancini6,hmfm}, t-J \cite{msm} and p-d \cite
{mancini2,mmvm,mmv} models.

The insufficiency of studying only those electronic states with energy close
to the Fermi level was shown in Ref.\ \cite{mancini2}. The reduced p-d
model, where the transitions to the lower Hubbard level are neglected, was
studied within a four-pole expansion for the Green's function. The fields
were taken as the bare $p$ electron, the upper Hubbard field, the $p$
electron field with spin-flips at copper sites and the $p$ electron operator
with p-d charge-transfer.

In this work we investigate the full p-d model with a
four-component field constituted by $p$ electrons, lower and upper
Hubbard operators for $d$ electrons and a fourth field which
describes $p$ electrons dressed by spin fluctuations of $d$
electrons. Some preliminary results were given in Ref.\
\cite{ManciniBarabanov}, where in particular we showed the
necessity of introducing the fourth field. The model treated in
terms of only the first three fields gives results not in
agreement with the numerical simulation data and that do not give
a satisfactory description of the experimental situation.

In Sec.\ \ref{sec:level1} we present the p-d Hamiltonian and introduce the
excitation modes represented in terms of composite field operators. The
Green's function for these operators is calculated in the four-pole
approximation and depends on a set of parameters expressed as equal-time
correlation functions. The study of these parameters is given in Sec.\ \ref
{sec:level2}, where by using the symmetry imposed by the Pauli principle we
derive a set of equations which allow us to obtain a complete
self-consistent solution.

The developing of numerical methods, as Quantum Monte Carlo (QMC)
and exact diagonalization \cite{dagotto}, offers a new challenge
to theoreticians. Before looking at the physical implication of
the model one should be confident with the solution by calculating
some quantities and comparing them with the data of numerical
simulations. According to this, in Sec.\ \ref{subsecA} we make a
detailed comparison of our results with the available data
obtained by QMC studies. The comparison shows a very good
agreement signaling us that the choice of the composite field and
the self-consistent procedure adopted are able to give a
reasonable solution of the model. We then proceed in sections
\ref{subsecB} and \ref{subsecC} to analyze some physical
properties. Section V is devoted to concluding remarks. Details of
calculations are given in Appendix.

\section{The p-d Model within the framework of Composite Operator Method}

\label{sec:level1}

In this section we consider the p-d model and we present the
relevant features of the Composite Operator Method. Starting from
the tight-binding model \cite{emery} composed of $d$ electrons in
$3d_{x^{2}-y^{2}}$ copper orbitals and $p$ electrons in $2p_{x,y}$
oxygen orbitals in a CuO$_{2}$ plane and considering only bonding
$p$ electrons \cite{Matsumotoxxx}, we keep the terms describing
the strong intra-atomic Coulomb repulsion at Cu sites and the $p$
and $d$ electrons hopping.

Therefore, the original Hamiltonian of the p-d model can be
written as

\begin{eqnarray}  \label{Hamiltonian}
H &=&\sum\limits_{i} \bigg[ (\varepsilon _{d}-\mu
)d^{\dagger}(i)d(i)+(\varepsilon _{p}-\mu )p^{\dagger}(i)p(i)
\nonumber \\ &+&  U n_{\uparrow }(i)n_{\downarrow }(i)\, +\, 2t\,
\left( p^{\gamma^{\dagger}}(i)d(i) \, +\, d^{\dagger}(i)p^{
\gamma}(i) \right) \bigg]
\end{eqnarray}

\noindent where in the spinor notation $p_{\sigma }^{\dagger }(i)$ and $%
d_{\sigma }^{\dagger }(i)$ create a $p$ electron and a $d$ electron on site $%
i\equiv ({\bf R}_{i},t_{i})$ with spin projection $\sigma \in
\{\uparrow ,\downarrow \}$, respectively. $\mu $ is the chemical
potential, the $U$ term describes the Coulomb repulsion at Cu
sites, $n_{\sigma }(i)=d_{\sigma }^{\dagger }(i)d_{\sigma }(i)$ is
the charge density operator of $d$ electrons with spin $\sigma $.
The $t$ term describes the p-d hopping and $p^{\gamma }(i)$ is
defined by

\begin{equation}
p^{\gamma}(i)=\sum\limits_{j}\gamma_{ij}p(j).
\end{equation}

\noindent Here $\gamma_{ij}$ is given by

\begin{equation}
\gamma_{ij}=\frac{a^{2}}{(2\pi )^{2}} \int_{\Omega_B} d^{2} {\bf k} \; e^{i
{\bf k}\cdot ({\bf R}_i - {\bf R}_j) } \; \gamma({\bf k}),
\end{equation}

\noindent where $\Omega _{B}$ is the volume of the first Brillouin zone, $%
\gamma({\bf k})=\sqrt{1-\alpha({\bf k})}$ and where for a two-dimensional
quadratic lattice with lattice constant $a$

\begin{equation}
\alpha({\bf k})=\frac{1}{2}\left[ \cos (k_{x}a)+\cos (k_{y}a)\right].
\end{equation}

Due to the strong correlation among electrons we introduce the
composite field

\begin{equation}
\Psi (i)\equiv \left(
\begin{array}{c}
p(i) \\
\xi (i) \\
\eta (i) \\
p_{s}(i)
\end{array}
\right),  \label{eq:fields}
\end{equation}

\noindent where

\begin{equation}
\xi(i)=\left[ 1-n(i)\right]d(i), \hskip 5mm
\eta(i)=n(i)d(i)
\end{equation}

\noindent are the Hubbard operators which describe the basic
excitations $ n\left( i\right) =0\leftrightarrow n\left( i\right)
=1$ and $n\left( i\right) =1\leftrightarrow n\left( i\right) =2$
on the lattice site $i$, respectively. The fourth field is chosen
as

\begin{equation}
p_{s}(i)=\sigma _{k}n_{k}(i)p^{\gamma }(i)-\frac{3c}{I_{22}}\xi
(i)-\frac{3b}{I_{33}}\eta (i).
\end{equation}

The parameters $b$ and $c$ and the quantities $I_{22}$ and
$I_{33}$ are given in Appendix.

We have introduced the charge- $\left( \mu =0\right) $ and spin-
$\left( \mu =1,2,3\right) $ density operator of $d$ electrons

\begin{equation}
n_{\mu }(i)=d^{\dagger }(i)\sigma _{\mu }d(i),
\end{equation}
and we are using the following notation

\begin{equation}
\sigma_{\mu }\equiv (1, \vec{\sigma}), \hskip 5mm
\sigma^{\mu}\equiv (-1, \vec{\sigma})
\end{equation}

\noindent with $\sigma _{k}$ $(k=1,2,3)$ being the Pauli matrices.

The choice of the composite field (\ref{eq:fields}) is dictated by
the following considerations. The strong intra-atomic Coulomb
interaction at Cu sites induces a splitting of the $d$ band into
the lower and the upper Hubbard subbands. The large covalence
between oxygen and copper electrons leads to a large fluctuation
of the energy of $p$ electrons at O sites. The field $p_{s}$,
describing $p-$ electronic excitations accompanied by the nearest
neighbor $d-$ electron spin fluctuations, represents electronic
excitations associated with the Cu-O bonds.

The Heisenberg equations of motion for the composite field $\Psi
\left( i\right) $ are

\begin{eqnarray}
i\frac{\partial }{\partial t}\Psi (i)=\left[ \Psi \left( i\right)
,H\right] \nonumber \\ =\left(
\begin{array}{l}
(\varepsilon _{p}-\mu )p(i)+2t\left[ \xi ^{\gamma }(i)+\eta ^{\gamma }(i)%
\right] \\ (\varepsilon _{d}-\mu )\xi (i)+2tp^{\gamma }(i)+2t\pi
(i) \\ (\varepsilon _{d}-\mu +U)\eta (i)-2t\pi (i) \\ (\varepsilon
_{p}-\mu )p_{s}(i)+\varepsilon _{pp}p^{\gamma }(i)+\varepsilon
_{p\xi }\xi (i) \\ \hskip5mm+\varepsilon _{p\eta }\eta
(i)+t_{p}\pi (i)+2t\kappa _{s}(i)
\end{array}
\right),  \label{eq:Heisenberg0}
\end{eqnarray}

\noindent where the coupling constants $\varepsilon _{pp}$, $\varepsilon
_{p\xi }$, $\varepsilon _{p\eta }$, $t_{p}$ and the higher order operators $%
\pi (i)$ and $\kappa _{s}(i)$ are defined by the following relations:

\begin{equation}
\varepsilon _{pp}=-\frac{6tc}{I_{22}},\hskip7mm\varepsilon _{p\xi }=\frac{%
3c(\varepsilon _{p}-\varepsilon _{d})}{I_{22}},
\end{equation}

\begin{equation}
\varepsilon _{p\eta }=\frac{3b(\varepsilon _{p}-\varepsilon _{\eta })}{I_{33}%
},\hskip7mmt_{p}=6t\left(
\frac{b}{I_{33}}-\frac{c}{I_{22}}\right),
\end{equation}

\begin{equation}
\pi (i)=\frac{1}{2}\sigma^{\mu }n_{\mu }(i)p^{\gamma }(i)+\xi (i)p^{\gamma
^{\dagger}}(i)\eta (i),
\end{equation}

\begin{eqnarray}
\kappa _{s}(i)= && \sigma_{k}d^{\dagger}(i)\sigma _{k}p^{\gamma
}(i)p^{\gamma }(i)  \nonumber \\
&& -\sigma _{k}p^{\gamma ^{\dagger}}(i)\sigma _{k}d(i)p^{\gamma }(i)
+\sigma_{k}n_{k}(i)d^{\gamma ^{2}}(i).
\end{eqnarray}

Let us introduce the thermal retarded Green's function

\begin{equation}
S(i,j)\equiv \left\langle R\left[ \Psi (i)\Psi ^{\dagger
}(j)\right] \right\rangle,  \label{eq:GreenF00}
\end{equation}
where $R$ is the usual retarded operator and the bracket $\left\langle
\cdots \right\rangle $ denotes the thermal average on the grand canonical
ensemble. Use of equation of motion (\ref{eq:Heisenberg0}) will generate an
infinite hierarchy of coupled equations and we need to introduce some
approximation. To this purpose, let us split the Heisenberg equation (\ref
{eq:Heisenberg0}) into a linear and nonlinear part

\begin{equation}
i\frac{\partial }{\partial t}\Psi (i)=\sum_{j}\varepsilon (i,j)\Psi
(j)+\delta J(i),
\end{equation}
where the equal-time correlation matrix $\varepsilon (i,j)$, the so-called
energy matrix, is fixed by the requirement that the nonlinear term $\delta
J(i)$ is orthogonal to the fundamental basis (\ref{eq:fields}):

\begin{equation}
\left\langle \{\delta J(i),\Psi ^{\dagger
}(j)\}_{E.T.}\right\rangle =0. \label{ortheq}
\end{equation}

In the framework of the pole approximation we neglect the nonlinear term $%
\delta J(i)$ in the equation of motion. Then, for a translational invariant
system the Fourier transform $S({\bf k},\omega )$ of the retarded propagator
has the following expression

\begin{equation}
S({\bf k},\omega )=\frac{1}{\omega -\varepsilon ({\bf k})}I({\bf k}),
\label{eq:GreenF0}
\end{equation}
where $\varepsilon ({\bf k})$, the Fourier transform of the energy matrix,
can be expressed as

\begin{equation}
\varepsilon ({\bf k})=m({\bf k})I^{-1}({\bf k}).
\end{equation}

In order to simplify the notation we introduced the so-called normalization
matrix $I({\bf k})$ and the so-called mass matrix $m({\bf k})$

\begin{equation}
I({\bf k})\equiv F.T.\left\langle \{\Psi (i),\Psi ^{\dagger
}(j)\}_{E.T.} \right\rangle ,  \label{eq:I}
\end{equation}

\begin{equation}
m({\bf k})\equiv F.T.\left\langle \{i\frac{\partial }{\partial
t}\Psi (i),\Psi ^{\dagger }(j)\}_{E.T.} \right\rangle,
\label{eq:m}
\end{equation}
where the symbol $F.T.$ denotes the Fourier transform.
Straightforward calculations give the expressions of these
matrices. The results for a paramagnetic state are given in
Appendix. The Green's function (\ref {eq:GreenF0}) can be put in
the spectral form

\begin{equation}
S({\bf k},\omega )=\sum\limits_{n=1}^{4}\frac{\sigma ^{(n)}({\bf k})}{\omega
-E_{n}({\bf k})+i\delta }.  \label{eq:GreenF}
\end{equation}

The energy-spectra \ $E_{n}({\bf k})$ \ are the eigenvalues of the $%
\varepsilon({\bf k})$ matrix and the spectral functions $\sigma^{(n)}({\bf k}%
)$ are given by

\begin{equation}
\sigma _{ab}^{(n)}({\bf k})=\sum_{c=1}^{4}\Lambda _{an}({\bf k})\,\Lambda
_{nc}^{-1}({\bf k})I_{cb}({\bf k}),
\end{equation}

\noindent where $a,\,b=1,\ldots ,4$ and where the columns of the $\Lambda (%
{\bf k})$ matrix are the eigenvectors of the $\varepsilon ({\bf k})$ matrix.

Summarizing, in our scheme of calculations firstly we choose a fundamental
basis of Heisenberg composite operators. The propagator for this basic field
is evaluated in a pole approximation, where the incoherent part is
neglected. The main ingredient of the calculation is the energy matrix. The
knowledge of this function will allow us to calculate the excitations of the
system and the spectral functions. The entire process requires a
self-consistent procedure which will be discussed in the next Section.

\section{Correlation functions and self-consistent equations}

\label{sec:level2}

In this section we introduce the correlation functions and give the
expressions of the self-consistent equations needed to calculate the
fermionic propagator.

As shown in Appendix, the Green's function depends on a set of parameters
expressed as static correlation functions of composite operators. Some of
these operators belong to the basic set (\ref{eq:fields}) and their
expectation values are directly connected to matrix elements of the Green's
function. Other operators are composite fields of higher order, out of the
basis (\ref{eq:fields}), and their correlation functions must be evaluated
by other means. This aspect is a general property of the Green's function
method \cite{mancini?}. These functions refer to a specific choice of the
Hilbert space and one must specify the proper representation. In particular,
the states must be constructed in such a way that the relations imposed by
the Pauli principle must be satisfied at level of expectation values.

Let us introduce the correlation function

\begin{equation}
C(i,j)= \left\langle \Psi(i) \Psi^{\dagger}(j) \right\rangle.
\end{equation}

By means of the spectral theorem and by recalling Eq.\
(\ref{eq:GreenF}), this function has the expression

\begin{eqnarray}
C(i,j) &=&\frac{a^{2}}{2(2\pi )^{2}}\sum\limits_{n=1}^{4}\int_{\Omega
_{B}}d^{2}ke^{i{\bf k}\cdot ({\bf R}_{i}-{\bf R}_{j})-iE_{n}({\bf k})\left(
t_{i}-t_{j}\right) }  \nonumber \\
&&\times \left[ 1+T_{n}({\bf k})\right] \sigma ^{(n)}({\bf k}),
\end{eqnarray}

\noindent where

\begin{equation}
T_{n}({\bf k})=\tanh \left( \frac{E_{n}({\bf k})}{2k_{B}T}\right).
\end{equation}

The parameters directly connected to the Green's function are: $n_d$, $a_s$,
$b$, $b_s$, $c$, $D$. By means of the definitions given in Appendix, we have
the self-consistent equations

\begin{equation}  \label{eq:SC1}
n_{d}=2(1-C_{22}-C_{33}),
\end{equation}

\begin{equation}  \label{eq:SC2}
a_{s}=C_{14}^{\gamma }+\frac{3c}{I_{22}}C_{12}^{\gamma }+\frac{3b}{I_{33}}%
C_{13}^{\gamma },
\end{equation}

\begin{equation}  \label{eq:SC3}
b=C_{13}^{\gamma },
\end{equation}

\begin{eqnarray}  \label{eq:SC4}
b_{s}=&& C_{24}^{\alpha }+C_{34}^{\alpha }+\frac{3c}{I_{22}}\left[
C_{22}^{\alpha }+C_{23}^{\alpha }\right]  \nonumber \\
&& +\frac{3b}{I_{33}}\left[ C_{23}^{\alpha }+C_{33}^{\alpha }\right],
\end{eqnarray}

\begin{equation}  \label{eq:SC5}
c=C_{12}^{\gamma },
\end{equation}

\begin{equation}  \label{eq:SC6}
D=I_{33}-C_{33}.
\end{equation}

We are using the following notation

\begin{equation}
C_{\mu \nu }=C_{\mu \nu }(i,i),
\end{equation}

\begin{equation}
C_{\mu \nu }^{\gamma }=\sum_{j}\gamma _{ij}C_{\mu \nu }(j,i)_{E.T.},
\end{equation}

\begin{equation}
C_{\mu \nu }^{\alpha }=\sum_{j}\alpha _{ij}C_{\mu \nu }(j,i)_{E.T.}
\end{equation}

with

\begin{equation}
\alpha_{ij}=\frac{a^{2}}{(2\pi )^{2}} \int_{\Omega_B} d^{2} {\bf k} \; e^{i
{\bf k}\cdot ({\bf R}_i - {\bf R}_j) } \; \alpha({\bf k}).
\end{equation}

The parameters not directly connected to the Green's function are $\mu $, $%
d_{s}$, $f$, $\chi _{s}$, defined in Appendix. In principle there are
different ways to calculate these parameters: decoupling approximation,
projection on the basis, use of linearized equations of motion. However, as
discussed in Refs. \cite{mancini6,mancini?}, there is only one definite way
to fix them: in order to have the right representation the Green's function
must satisfy the equation

\begin{equation}
\lim_{j\to i^{-}}S(i,j)=\left\langle \Psi (i)\Psi
^{^{\dagger}}(i)\right\rangle,
\end{equation}

\noindent where the r.h.s. is calculated by means of the Pauli principle.

Use of this equation leads to the following self-consistent equations

\begin{equation}  \label{eq:SC7}
C_{23}=0,
\end{equation}

\begin{equation}  \label{eq:SC8}
C_{24}=3C_{12}^{\gamma }-\frac{3C_{12}^{\gamma }C_{22}}{I_{22}},
\end{equation}

\begin{equation}  \label{eq:SC9}
C_{34}=-\frac{3C_{13}^{\gamma }C_{33}}{I_{33}},
\end{equation}

\begin{equation}  \label{eq:SC10}
n_{T}=4-2(C_{11}+C_{22}+C_{33}),
\end{equation}

\noindent where $n_{T}=n_{p}+n_{d}$ is the total particle number
and $n_{p}$ is the number of $p$ electrons. Equations
(\ref{eq:SC1})-(\ref{eq:SC6}) and (\ref{eq:SC7})-(\ref{eq:SC10})
constitute a set of coupled equations which will fix the
parameters in a self-consistent way.

Summarizing, we have ten parameters and ten self-consistency
equations which allow us to compute the Green's function and the
properties of the system in a fully self-consistent way.

Results of calculations are presented in the next Section. As as a
comparison we also present results in the case where we do not
take into account the Pauli principle and we express the
parameters $d_{s},f,\chi _{s}$ by means of the following
decoupling equations

\begin{equation}
f\approx -C_{11}^{\gamma \gamma }\left[ C_{12}^{\gamma }+C_{13}^{\gamma }%
\right] ,  \label{eq:decf}
\end{equation}

\begin{equation}
d_{s}\approx \left[ C_{12}^{\gamma \alpha }+C_{13}^{\gamma \alpha }\right] %
\left[ C_{22}^{\alpha }+2C_{23}^{\alpha }+C_{33}^{\alpha }\right]
, \label{eq:decd}
\end{equation}

\begin{equation}
\chi _{s}\approx -2\left[ C_{22}^{\alpha }+2C_{23}^{\alpha }+C_{33}^{\alpha }%
\right] ^{2}.  \label{eq:decchi}
\end{equation}

It should be noted that for certain values of the external
parameters some instabilities appear in the iteration procedure
and no solution of the self-consistent equations is found. This
aspect is not present when the decoupling scheme
(\ref{eq:decf})-(\ref{eq:decchi}) is used and may be related to
the fact that more asymptotic fields are necessary in order to
have a proper representation for the Green's function
\cite{mancini?}.

\section{Results}

\label{sec:level3}

This Section is organized as follows. The first part is devoted to
compare the results of our calculations with the data of numerical
analysis by Quantum Monte Carlo and exact diagonalization. As it
will be shown in Sec.~\ref{subsecA}, for all local quantities the
agreement between numerical and COM results is excellent. Once we
are confident to have a reasonable solution of the model, we go to
the next step where we study some physical properties in order to
verify if the p-d model is a realistic model for cuprate
superconductors. This is done in the second part, Sections \ref
{subsecB} and \ref{subsecC}, where the density of states and Fermi
surface are analyzed. In this study the values of the model
parameters have been taken according to the results suggested by
{\it ab initio} calculation \cite {McMahan}: $t=1eV$, $U=6eV$,
$\Delta =2eV$, where $\Delta =\varepsilon _{d}+U-\varepsilon _{p}$
is the charge-transfer energy.

In this work all energy are given in units of $t$ and measured with respect
to the atomic level $\varepsilon_p =0$.

\subsection{Comparison with QMC data}

\label{subsecA}

Here we compare COM results with QMC calculations and Lanczos
diagonalization results \cite{muramatsu1}, \cite{Scalettar}, \cite
{muramatsu2}.

We introduce the squared local magnetic moment for $d$ electrons as

\begin{equation}
S_{z}^{2}={\frac{1}{N}}\sum_{i}\left\langle {\left[ n_{d\uparrow
}(i)-n_{d\downarrow }(i)\right] }^{2}\right\rangle .  \label{eq:localmoment}
\end{equation}

This quantity can be expressed for paramagnetic case through the double
occupancy and the number of $d$ electrons as follows: $S_{z}^{2}=n_{d}-2D$.

Following Ref.\ \cite{muramatsu1} we can distinguish two different
regimes according to the values of the $U$ Hubbard repulsion and
the charge-transfer energy $\Delta $. In the region $U>\Delta $
the insulating properties of the system are characterized by a
charge-transfer gap. Instead, in the region $U<\Delta $ the system
is in a Mott-Hubbard regime. Both regions are studied in the
following.

\begin{figure}[h!]
\begin{center}
\includegraphics*[width=8cm]{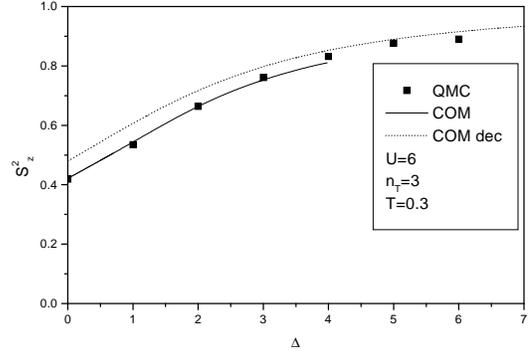}
\end{center}
\caption{The squared local magnetic moment $S_z^2$ against the parameter $%
\Delta$; the values of external parameters $U$, $n_{T}$ and $T$
are given in the figure. The solid line is COM prediction with
Pauli principle, the dotted one with decoupling. The squares are
QMC data \protect\cite {muramatsu1}.} \label{sqdeltam}
\end{figure}

\begin{figure}[h!]
\begin{center}
\includegraphics*[width=8cm]{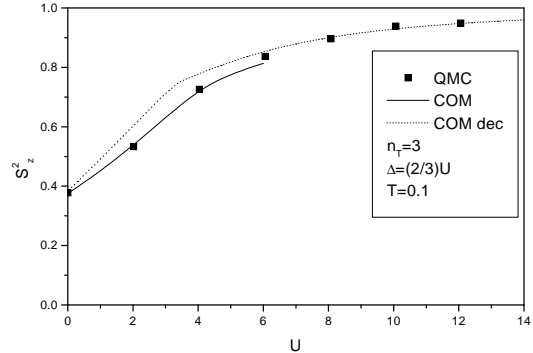}
\end{center}
\caption{The squared local magnetic moment $S_z^2$ against the parameter $U$%
; the values of external parameters $n_{T}$, $\Delta$ and $T$ are
given in the figure. The solid line is COM prediction with Pauli
principle, the dotted one with decoupling. The squares are QMC
data \protect\cite {muramatsu1}.} \label{squm}
\end{figure}

In Fig.~\ref{sqdeltam} and Fig.~\ref{squm} we plot the squared local
magnetic moment against the parameters $\Delta$ and $U$, respectively. $%
S_z^2 $ is an increasing function of of these parameters. Figure
~\ref{sqdeltam} shows that the local magnetic moment $S_z^2$ takes
the smallest value when $\Delta$ approaches zero; in such case $p$
and $d$ upper Hubbard levels coincide and due to strong mixing the
double occupancy takes a large value.

Instead, when $\Delta$ becomes larger than $U$ the system changes from
charge-transfer insulator to Mott-Hubbard insulator and the local magnetic
moment $S_z^2$ becomes independent on $\Delta$. In this regime we are
addressing a single-band 2D Hubbard model solution.

An increasing behaviour of $S_z^2$ is also possible for
charge-transfer insulator as it is shown in Fig.~\ref{squm}, where
$\Delta$ is taken equal to $2U/3$. In such case double occupancy
of $d$ electrons decreases with increasing distance between bands
and the local magnetic moment saturates.

\begin{figure}[h!]
\begin{center}
\includegraphics*[width=8cm]{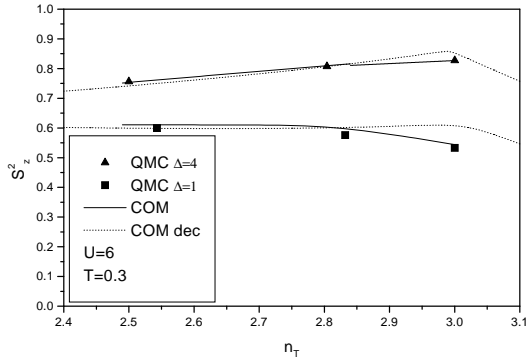}
\end{center}
\caption{The squared local magnetic moment $S_z^2$ against $n_{T}$ for $%
\Delta=1$ and $\Delta=4$; the values of external parameters $U$
and $T$ are given in the figure. The solid line is COM prediction
with Pauli principle, the dotted one with decoupling. The squares
and the triangles are QMC data \protect\cite{muramatsu1}.}
\label{sqntm}
\end{figure}

\begin{figure}[h!]
\begin{center}
\includegraphics*[width=8cm]{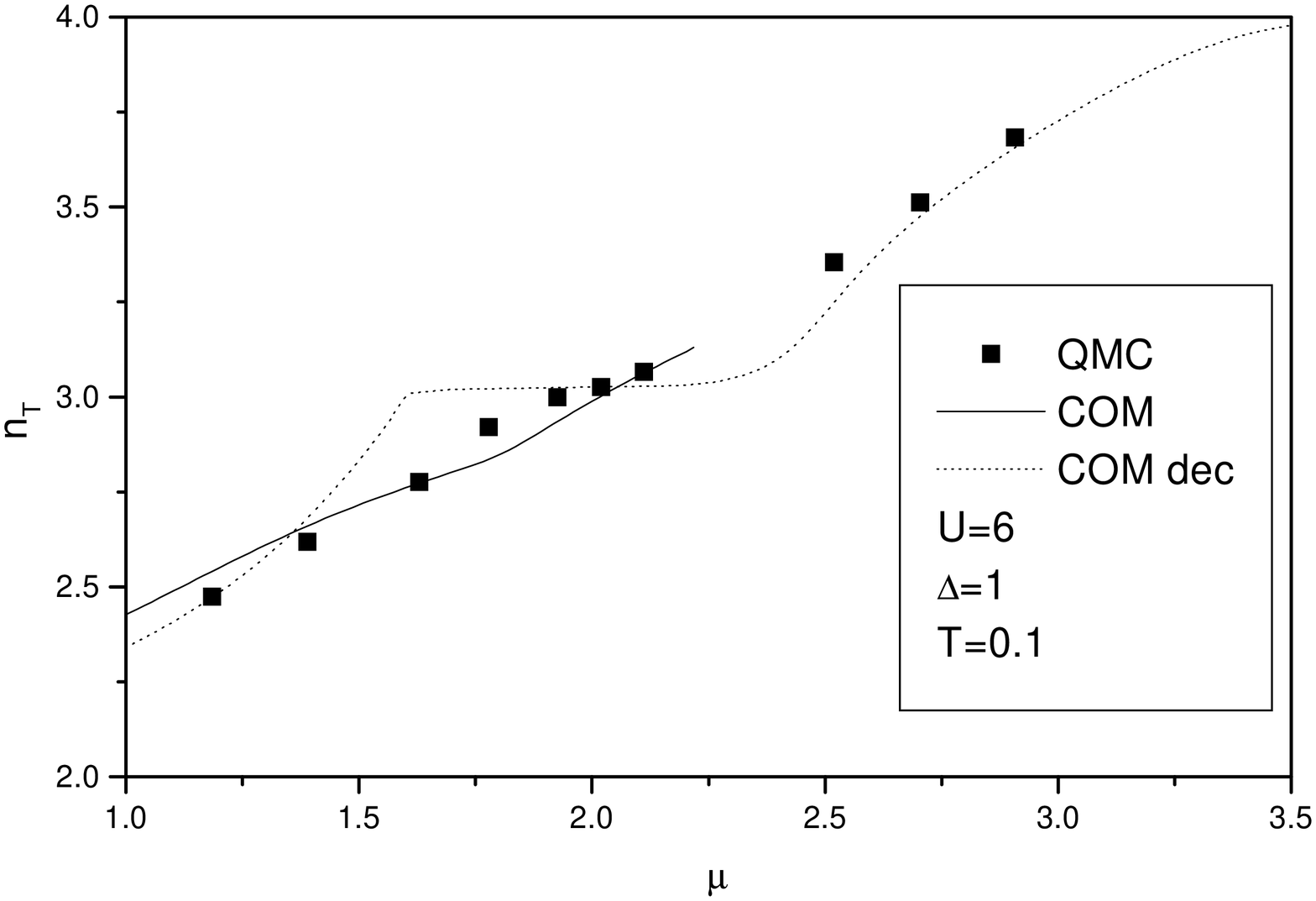}
\end{center}
\caption{The $n_{T}$-$\protect\mu$ plot for the values of external
parameters $U=6$, $\Delta=1$ and $T=0.1$. The squares are QMC data
\protect\cite{muramatsu1}, the solid line is COM prediction with
Pauli principle, the dotted one is with decoupling.} \label{ntmum}
\end{figure}

The dependence of the squared local magnetic moment on the total
number of particles $n_T$ for $\Delta=1$ and $\Delta=4$ and for
the values of external parameters $U=6$, $T=0.3$ is shown in
Fig.~\ref{sqntm}.

We also report results for the case when the parameters $d_{s},f,\chi _{s}$
are expressed by means of decoupling equations (\ref{eq:decf})-(\ref
{eq:decchi}) (dotted line).

In all cases the agreement with QMC results is very good,
specially in the case when we take into account the Pauli
principle.

From Fig.~\ref{sqntm} we see that the local magnetic moment only
slightly changes with doping. Such fact was also observed in
neutron scattering experiments on $La_{2-x}Sr_xCuO_4$
\cite{Birgeneau}, where it was stated that doping destroys
antiferromagnetic spin correlations but does not destroy local
magnetic moments.

\begin{figure}[h!]
\begin{center}
\includegraphics*[width=8cm]{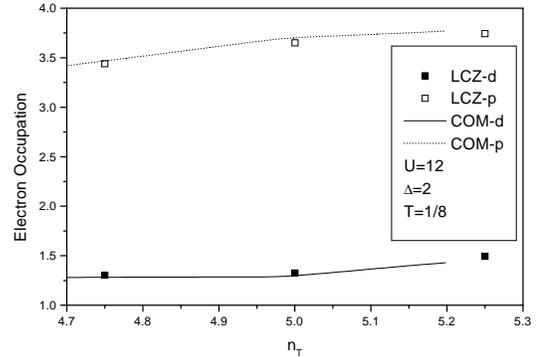}
\end{center}
\caption{$n_{d}$ and $n_{p}$ against the total electron occupation $%
n_{T}=n_{d}+2n_{p}$ for the values of external parameters $U=12$,
$\Delta=2$ and $T=1/8$. The squares are Lanczos diagonalization
results \protect\cite {Scalettar}, the solid and dotted lines are
COM calculation.} \label{eont12s}
\end{figure}

\begin{figure}[h!]
\begin{center}
\includegraphics*[width=8cm]{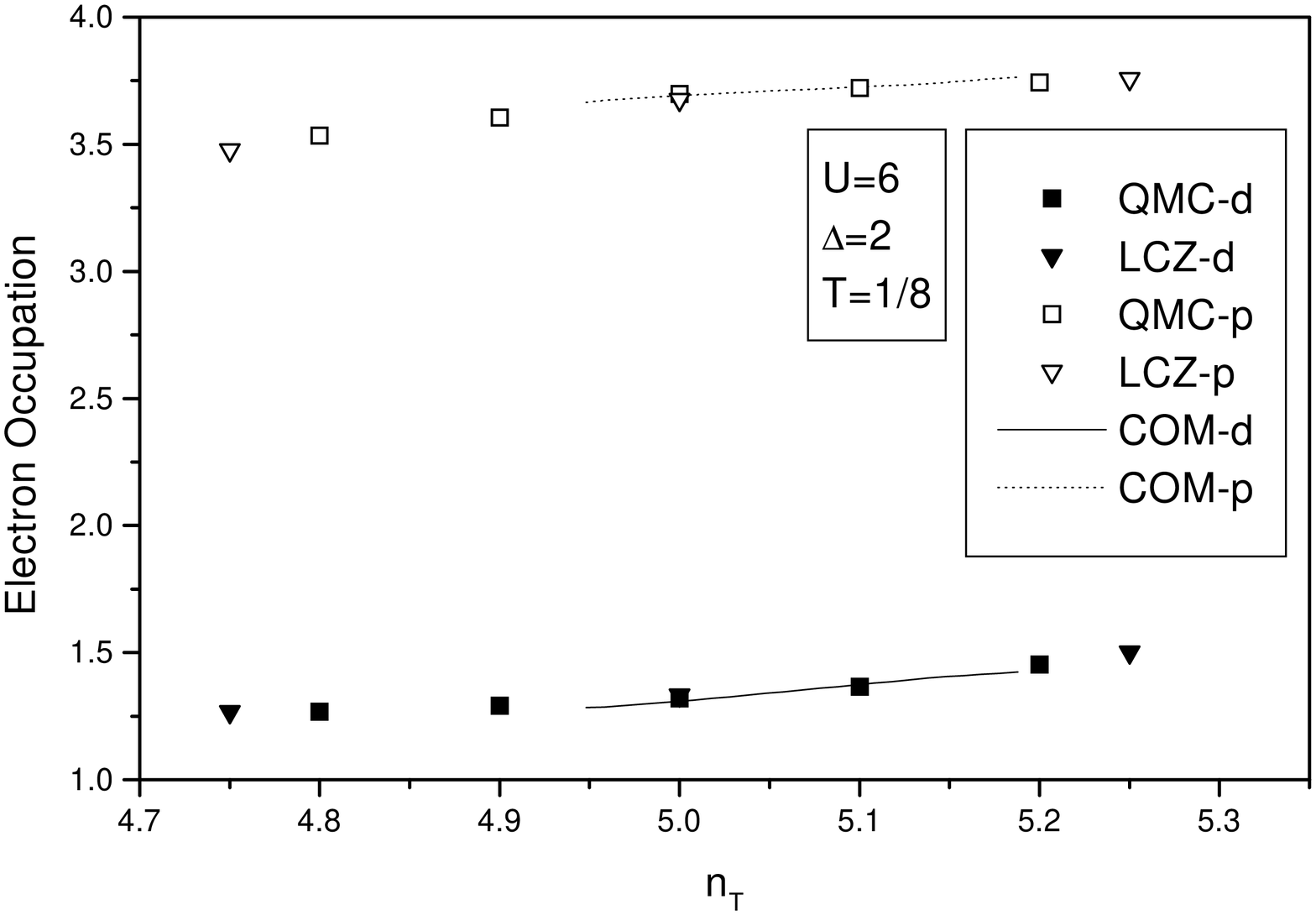}
\end{center}
\caption{$n_{d}$ and $n_{p}$ against the total electron occupation $%
n_{T}=n_{d}+2n_{p}$ for the values of external parameters $U=6$,
$\Delta=2$ and $T=1/8$. The squares and the triangles are QMC and
Lanczos diagonalization results, respectively
\protect\cite{Scalettar}; the solid and dotted lines are COM
predictions.} \label{eont6s}
\end{figure}

In a previous paper \cite{mmvm} the transition between $p$ level
and the lower Hubbard level $\xi $ was not taken into account, and
the the number of $d$ electrons $n_{d}$ was calculated only
approximately. As a result the agreement with QMC data was only
qualitative.

In Fig.~\ref{ntmum} the dependence of $n_{T}$ on the chemical
potential is drawn for $U=6$, $\Delta =1$ and $T=0.1$; a good
agreement between COM predictions and QMC results is found. As a
comparison, it is also reported the case when the Pauli principle
is not taken into account; in this case the agreement with QMC
simulations is not so good.

In Fig.~\ref{eont12s} and Fig.~\ref{eont6s} we present also a very
good agreement of our results with Lanczos diagonalization and QMC
calculations given in Ref.\ \cite{Scalettar}. The authors of this
work write the full number of holes as $n=n_{Cu}+2n_{O}$, where
the double number of oxygen holes is due to the two oxygen ions in
the cell. We remind that in the p-d model there are two bonding
and two nonbonding electrons per site. To make a comparison with
results of Ref.\ \cite{Scalettar} we have to take into account the
Fermi occupation number of these two unbonding orbitals and to
rewrite equation (\ref{eq:SC10}) as

\begin{eqnarray}  \label{eq:SC10a}
n_{T} & = & 4-2(C_{11}+C_{22}+C_{33})  \nonumber \\ & + &
{\frac{2}{{\exp((\varepsilon_p - \mu)/T) + 1}}}.
\end{eqnarray}

For the values of external parameters given in Fig.~\ref{eont12s}
and Fig.~\ref{eont6s} the system is in the charge-transfer regime
and in the vicinity of $n_{T}=1$ holes prefer to go to $p$
orbitals and electrons to $d$ orbitals.

\begin{figure}[h!]
\begin{center}
\includegraphics*[width=8cm]{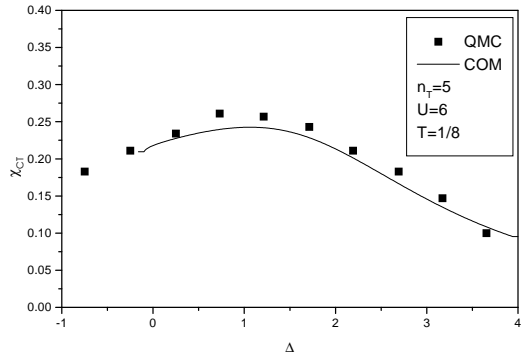}
\end{center}
\caption{The charge-transfer susceptibility $\protect\chi_{CT}$ at ${\bf q}%
=0 $ versus $\Delta$ for the values of external parameters $U=6$,
$n_{T}=5$ and $T=1/8$; the squares are QMC results
\protect\cite{Scalettar} and the solid line is COM prediction.}
\label{ctdeltas}
\end{figure}

In Fig.~\ref{ctdeltas} it is shown the dependence of the
charge-transfer susceptibility at ${\bf q} = 0$ on charge-transfer
energy $\Delta$. This quantity \cite{Scalettar} is given by

\begin{equation}  \label{eq:CT}
\lim_{{\bf q}\to 0}\chi_{CT}({\bf q})= {\frac{{\partial}}{{\partial \Delta}}}
\langle n_d - 2 n_p \rangle.
\end{equation}

The agreement between COM predictions and QMC simulations is very
good.

\begin{figure}[h!]
\begin{center}
\includegraphics*[width=8cm]{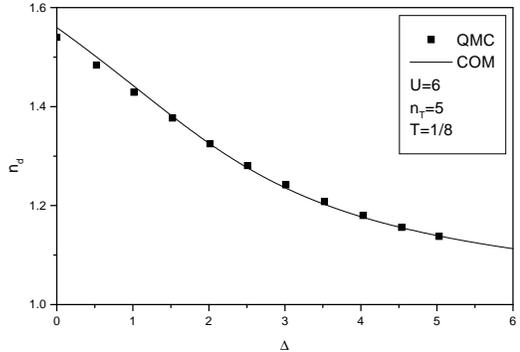}
\end{center}
\caption{The $n_{d}$-$\Delta$ plot for the values of external parameters $%
U=6 $, $n_{T}=5$ and $T=1/8$; the squares are QMC results
\protect\cite {Scalettar} and the solid line is COM prediction.}
\label{nddeltas}
\end{figure}

\begin{figure}[h!]
\begin{center}
\includegraphics*[width=8cm]{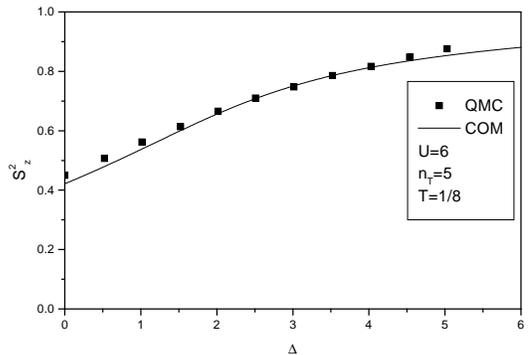}
\end{center}
\caption{The squared magnetic moment $S_{z}^{2}$ against $\Delta$
for the
values of external parameters $U=6$, $n_{T}=5$ and $T=1/8$; the squares are $%
QMC$ results \protect\cite{Scalettar} and the solid line is COM
result.} \label{sqdeltas}
\end{figure}

\begin{figure}[h!]
\begin{center}
\includegraphics*[width=8cm]{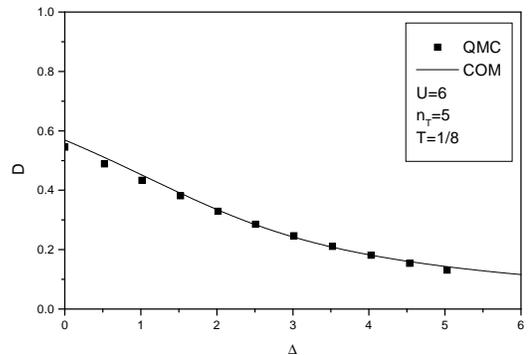}
\end{center}
\caption{The double occupancy $D$ versus $\Delta$ for the values
of external parameters $U=6$, $n_{T}=5$ and $T=1/8$; the squares
are QMC results \protect\cite{Scalettar} and the solid line is COM
prediction.} \label{ddeltas}
\end{figure}

In Fig.~\ref{nddeltas}, Fig.~\ref{sqdeltas} and Fig.~\ref{ddeltas}
we plot the quantities $n_d$, $S^{2}_{z}$ and $D$ against the
charge-transfer gap $ \Delta$ for the values of external
parameters $U=6$, $n_T=5$ and $T=1/8$. The agreement between COM
prediction and QMC simulation \cite{Scalettar} is excellent. The
dependence of the calculated quantities on external parameters has
the same behaviour as in Fig.~\ref{sqdeltam}. We want to note once
more that the holes introduced in CuO$_{2}$ planes reside
primarily on oxygen sites, as it is well-known experimentally.

\begin{figure}[h!]
\begin{center}
\includegraphics*[width=8cm]{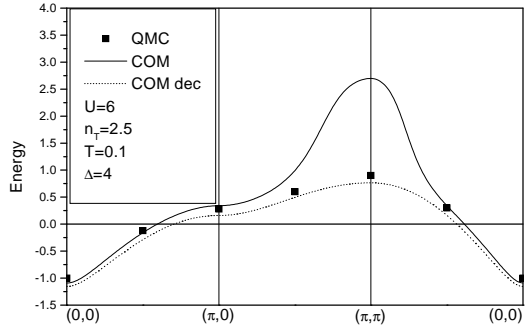}
\end{center}
\caption{The band structure for the Zhang-Rice singlet for
$n_{T}=2.5$. The squares are QMC data \protect\cite{muramatsu2},
the solid line is COM prediction with Pauli principle, the dotted
one is with decoupling.} \label{zr25}
\end{figure}

\begin{figure}[h!]
\begin{center}
\includegraphics*[width=8cm]{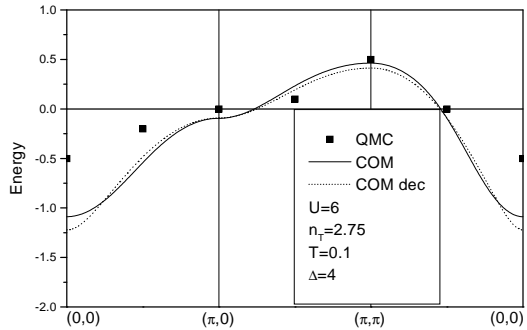}
\end{center}
\caption{The band structure for the Zhang-Rice singlet for
$n_{T}=2.75$. The squares are QMC data \protect\cite{muramatsu2},
the solid line is COM prediction with Pauli principle, the dotted
one is with decoupling.} \label{zr275}
\end{figure}

In Fig.~\ref{zr25} and Fig.~\ref{zr275} the band dispersion of
Zhang-Rice singlet for $n_{T}=2.5$ and $n_{T}=2.75$ is shown. The
squares are QMC data from \cite{muramatsu2} and the solid line is
the COM result. The data differ significantly only for the values
of dispersion energies far from the chemical potential.

Summarizing, in this section we have presented a detailed
comparison of COM results with the available data by numerical
simulation. It is worth noticing that the formulation is fully
self-consistent and no adjustable parameter is used.

\subsection{Density of states}

\label{subsecB}

In this paragraph and in the next one we present an analysis of the band
structures of the model and we give a description of the relevant physics
near the Fermi level.

The density of states (DOS) for $p$ and $d$ electrons is
respectively given by the following expressions

\begin{equation}  \label{eq:pDOS}
N_{p}(\omega ) = \frac{ a^2 }{(2\pi)^{2}} \int\limits_{\Omega _{B}}d^{2}{\bf %
k} \sum\limits_{n=1}^{4} \sigma_{11}^{(n)}({\bf k} ) \delta \left( \omega
-E_{n}({\bf k})\right),
\end{equation}

\begin{eqnarray}  \label{eq:dDOS}
N_{d}(\omega ) =&& \frac{ a^2 }{(2\pi)^{2}} \int\limits_{\Omega _{B}}d^{2}%
{\bf k} \sum\limits_{n =1}^{4}[ \sigma_{22}^{(n)}({\bf k}) +
2\sigma_{23}^{(n)}({\bf k})  \nonumber \\
&& + \sigma_{33}^{(n)}({\bf k})] \delta (\omega -E_{n}({\bf k})),
\end{eqnarray}

\noindent where the spectral functions $\sigma^{(n)}_{\alpha \beta} ({\bf k}%
) $ and the energy bands $E_{n}({\bf k})$ can be calculated as
indicated in Sec.\ \ref {sec:level1}.

\begin{figure}[h!]
\begin{center}
\includegraphics*[width=8cm]{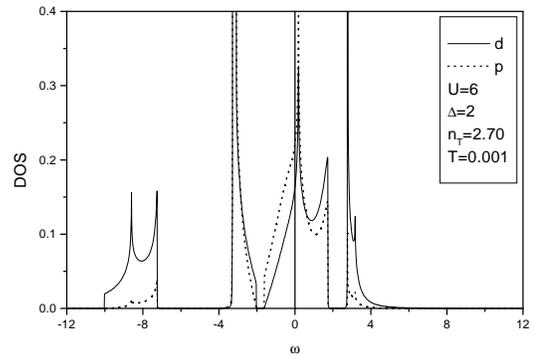}
\end{center}
\caption{$p-DOS$ (dotted line) and $d-DOS$ (solid line) for
$n_{T}=2.70$. With respect to the chemical potential
$\protect\varepsilon_{p}=-1.613$.} \label{dos27}
\end{figure}

\begin{figure}[h!]
\begin{center}
\includegraphics*[width=8cm]{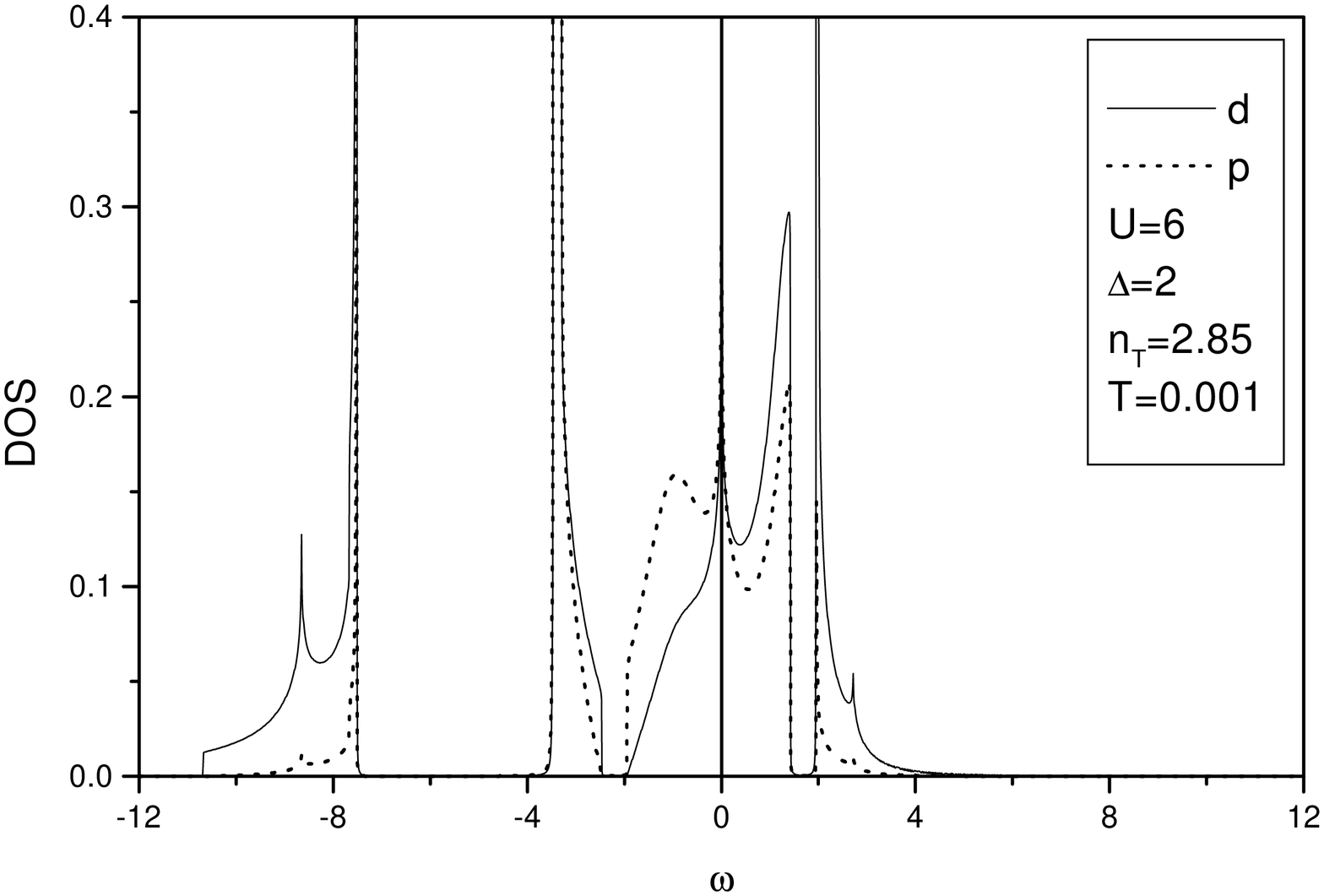}
\end{center}
\caption{$p-DOS$ (dotted line) and $d-DOS$ (solid line) for
$n_{T}=2.85$. With respect to the chemical potential
$\protect\varepsilon_{p}=-1.945$.} \label{dos285}
\end{figure}

\begin{figure}[h!]
\begin{center}
\includegraphics*[width=8cm]{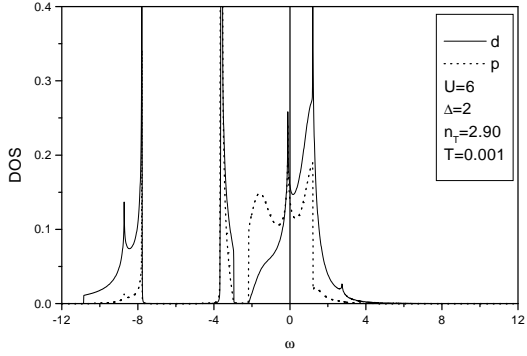}
\end{center}
\caption{$p-DOS$ (dotted line) and $d-DOS$ (solid line) for
$n_{T}=2.90$. With respect to the chemical potential
$\protect\varepsilon_{p}=-2.174$.} \label{dos29}
\end{figure}

In Figs.~\ref{dos27}, \ref{dos285} and \ref{dos29} we present the
DOS of $d$ electrons (solid line) and $p$ electrons (dotted lines)
for $ U=6 $, $\Delta =2$, $T=0.001$ and for $n_T=2.70$, $2.85$ and
$2.90$, respectively. The solid vertical line at $\omega =0$
indicates the chemical potential.

As it can be seen from these pictures there are four bands. The
lower Hubbard band located at $\omega \approx -9$ represents the
$\xi$ excitations; this band is filled mainly by $d$ electrons and
it has a small fraction of $p$ electrons. The upper Hubbard band,
located at $ \omega\approx 3$, comes from $\eta$ operator
excitations; it has mainly the weight of $d$ electrons. The band
coming from the $\varepsilon_p$ atomic level is located around
$\omega \approx -3$. It is filled by $p$ electrons, but due to
strong mixing it also contains a large fraction of $d$ electrons.
Finally, the Zhang-Rice singlet band is situated around the
chemical potential; this band has almost equal fraction of $p$ and
$d$ electrons.

We want to stress the importance of studying the Zhang-Rice
excitation as independent field in the initial set
(\ref{eq:fields}). As shown in Ref.\ \cite{ManciniBarabanov}, if
one does not initially consider the $p_s$ excitation, the
information about $p$ and $d$ electrons coupling is lost after
taking averages (\ref{eq:I}), (\ref{eq:m}) and one does not get
the band dispersion situated at the chemical potential when the
total number of particles is $n_{T}\approx 3$.

The singlet coupling of $p$ and $d$ electrons was investigated by
means of COM in the earlier work of Matsumoto {\it et al.}
together with one of us \cite{mancini1}. They obtained the singlet
excitation band at the chemical potential and studied how the
density of states depends on the variation of the correlators.

In this work we considered the full p-d model, as described by
Eq.\ (\ref{Hamiltonian}), and we used additional equations
supplied by Pauli principle (Eqs.\ (\ref{eq:SC7}) -(\ref{eq:SC9}))
to calculate the values of these correlators.

\subsection{An analysis of the experimental data}

\label{subsecC}

We investigated the region of total number of electrons around
$n_T=3$, which corresponds to the initial situation where only one
electron is in the copper $d$ orbital due to the strong Hubbard
repulsion and two electrons in the bounded $p$ orbitals of oxygen.

\begin{figure}[h!]
\begin{center}
\includegraphics*[width=8cm]{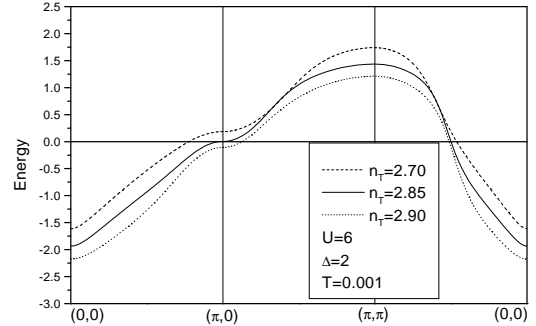}
\end{center}
\caption{The band structure for the Zhang-Rice singlet for $%
n_{T}=2.70,2.85,2.90$ and for $U=6$, $\Delta=2$ and $T=0.001$.}
\label{zr0001}
\end{figure}

In Fig.~\ref{zr0001} the band structure for the Zhang-Rice singlet
is drawn for the values of external parameters given in the
pictures; we see that at the momentum ${\bf k}=(\pi, 0)$ such band
has a saddle-point. This saddle-point leads to the van-Hove
singularity in the density of states (see Fig.~\ref{dos27},
Fig.~\ref{dos285} and Fig.~\ref{dos29}) and it was observed in
experiments \cite{tobin,abrikos,gofron}.

We note that for $U=6$ and $\Delta=2$ the coincidence of the
chemical potential and van-Hove singularity takes place at the
value of total number of particles $n_T=2.85$, corresponding to
the hole doping $\delta =0.15$. This results to an enhancement of
the thermodynamical properties such as specific heat and spin
magnetic susceptibility, as observed in the two-dimensional
Hubbard model \cite{mancini1,mancini5,mancinisusceptibility}. Upon
increasing the particle number up to $n_T=2.9$ the gap between the
upper Hubbard band and the Zhang-Rice singlet band disappears.

\begin{figure}[h!]
\begin{center}
\includegraphics*[width=7cm]{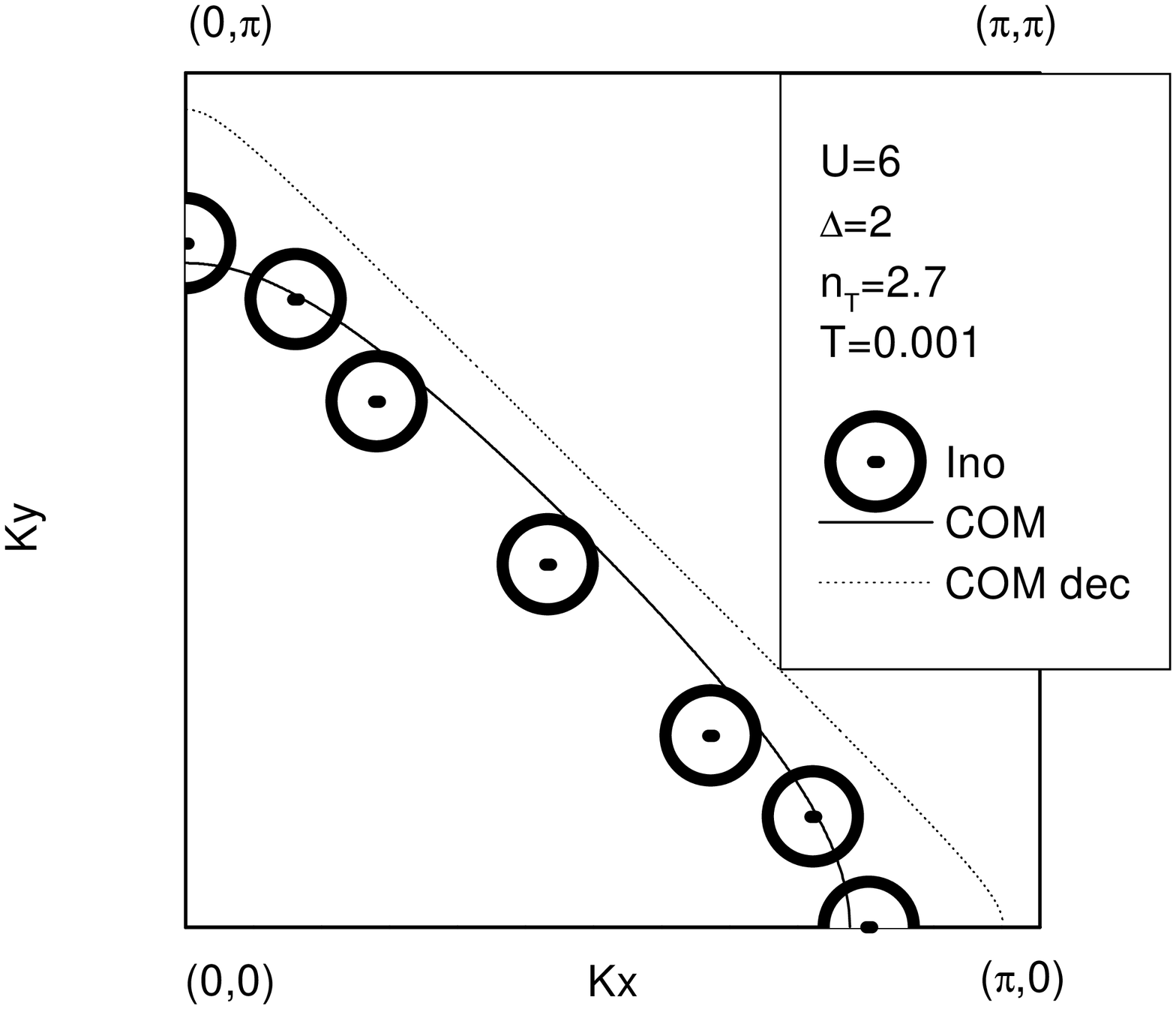}
\end{center}
\caption{The Fermi Surface for $U=6$, $\Delta=2$, $n_{T}=2.7$ and
$T=0.001$; the solid line is COM prediction with Pauli principle,
the dotted one with decoupling. The circles are experimental
results \protect\cite{ino}.} \label{e27}
\end{figure}

\begin{figure}[h!]
\begin{center}
\includegraphics*[width=7cm]{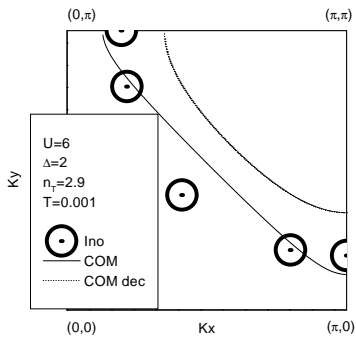}
\end{center}
\caption{The Fermi Surface for $U=6$, $\Delta=2$, $n_{T}=2.9$ and
$T=0.001$; the solid line is COM prediction with Pauli principle,
the dotted one with decoupling. The circles are experimental
results \protect\cite{ino}.} \label{e29}
\end{figure}

In Fig.~\ref{e27} and Fig.~\ref{e29} we plot the Fermi Surface for
the values of the external parameters $U=6$, $\Delta =2$ and
$T=0.001$ and for total number of particles $n_T=2.7$ and
$n_T=2.9$ respectively. The values of the temperature and of the
total number of particles were chosen in order to compare the
Fermi surface calculated by means of COM with the experimental one
measured by Ino {\it et al} \cite{ino} and shown by the circles in
the pictures.

We see that at the total number of particles $n_T=2.85$, when the
chemical potential crosses the van Hove singularity, the Fermi
surface changes its shape from the electron-like in the overdoped
regime ($n_T=2.7$) to the hole-like in the underdoped regime
($n_T=2.9$). This is in agreement with ARPES results \cite{ino}
and Hall coefficient measurements \cite{hall}.

As a comparison, we also plot the Fermi Surface when we do not take into
account the Pauli principle (dotted line); the agreement with experimental
results is not good in this case.


\section{Conclusions}

\label{sec:level4}

The p-d model was studied in order to describe energetic
properties of charge excitations in the CuO$_{2}$ planes of
cuprates in the normal state. Excitations are described in terms
of composite operators in the CuO$_{2}$ cluster. The most
important excitation appears to be the Zhang-Rice singlet. The
oxygen hole doped into the CuO$_{2}$ plane couples with Cu
electron forming a singlet state. The band of these states is
located between the oxygen level and the copper upper Hubbard
atomic level. By hole doping the Fermi level crosses this band and
at definite value of doping it coincides with a van-Hove
singularity in the density of states. This singularity is formed
by the saddled shape of the band dispersion at the Brillouin zone
points $(\pi,0)$ and $(0,\pi)$ as it was also obtained in
experiments and in QMC calculations.

The large volume of the Fermi surface at low doping results from
the small spectral weight of the field describing $p$ electrons
dressed by spin-flips of $d$ electrons. Such reduction of the
spectral weight is possible due to its redistribution to the other
bands strongly mixed with Zhang-Rice singlet band.

By using a four-pole approximation in the framework of the Composite
Operator Method various local quantities have been calculated as functions
of model and physical parameters. The results of calculations have been
presented in Sec.\ \ref{subsecA} and compared with the results of numerical
analysis.

We want to stress the relevance of Pauli principle in treating
with strongly correlated electronic systems within our method. The
comparison of COM results with numerical data, concerning local
properties and band structures, and experimental measurements
obtained by Ino {\it et al} \cite {ino} looks to have a good
agreement when we take into account the Pauli principle in
computing the fermionic propagator.

As last remark, we want to note that we can calculate the two-particle
Green's functions in the one-loop approximation \cite{mancini1} by using the
single-particle propagator. Calculations in this direction are now in
progress.

\acknowledgments We wish to thank Dr. A. Avella for fruitful
discussions and useful comments.


\appendix

\section*{The matrices $I$ and $m$ and the self-consistency parameters.}

\label{app.A}

In this appendix we give the explicit form of matrices $I$ and $m$ and we
also report the expression of the self-consistency parameters.

The $I$ matrix is diagonal and its elements are: 

\begin{equation}
\begin{array}{ccc}
I_{11}=1, & I_{22}=1-n_{d}/2, & I_{33}=n_{d}/2,
\end{array}
\end{equation}

\begin{eqnarray}
I_{44} &=&3(n_{d}-2D-\alpha ({\bf k})\chi _{s})+4a_{s}  \nonumber \\
&&-9(\frac{c^{2}}{I_{22}}+\frac{b^{2}}{I_{33}}).
\end{eqnarray}

The quantities $n_{d}$ and $D$ are the particle number and the
double occupancy per site of $d$ electrons, respectively:

\begin{equation}
n_{d}=\left\langle d^{\dagger }\left( i\right) d\left( i\right)
\right\rangle ,
\end{equation}

\begin{equation}
D=\left\langle n_{\uparrow }\left( i\right) n_{\downarrow }\left( i\right)
\right\rangle .
\end{equation}

The other parameters are defined by

\begin{equation}
\begin{array}{cc}
b=\left\langle p^{\gamma }(i)\eta ^{\dagger }(i)\right\rangle , &
c=\left\langle p^{\gamma }(i)\xi ^{\dagger }(i)\right\rangle ,
\end{array}
\end{equation}

\begin{equation}
\begin{array}{cc}
a_{s}=\left\langle p^{\gamma }(i)p_{s}^{\dagger }(i)\right\rangle , & \chi
_{s}=\frac{1}{3}\left\langle n_{k}(i)n_{k}^{\alpha }(i)\right\rangle .
\end{array}
\end{equation}

The $m$ matrix is hermitician and its elements have the expressions:

\begin{equation}
\begin{array}{ll}
m_{11}=\varepsilon _{p}-\mu, & m_{12}=2tI_{22}\gamma ({\bf k}), \\
m_{13}=2tI_{33}\gamma ({\bf k}), & m_{14}=0, \\
m_{22}=(\varepsilon _{d}-\mu )I_{22}+2t(c-b), & m_{23}=2(b-c), \\
m_{24}=2tI_{\pi p_{s}}, & m_{34}=-m_{24}, \nonumber
\end{array}
\end{equation}

\begin{equation}
\begin{array}{l}
m_{33}=(\varepsilon _{d}+U- \mu )I_{33}-2t(b-c), \\
m_{44}=(\varepsilon _{p}-\mu )I_{44}+2tI_{k_{s}p_{s}}+t_{p}I_{\pi {p}_{s}},
\end{array}
\end{equation}

\noindent where

\begin{equation}
I_{\pi {p}_{s}}=\frac{3}{2}(n_d-2D-\alpha({\bf k}) \chi_s) + 2a_s -\frac{%
t_p(b-c)}{2t},
\end{equation}

\begin{equation}
I_{\kappa_{s}\xi}= \frac{1}{2t}(2tI_{\pi p_{s}}-\varepsilon_{p\xi
}I_{22}-t_{p}(c-b)),
\end{equation}

\begin{equation}
I_{\kappa_{s}\eta}=-\frac{1}{2t}(2tI_{\pi p_{s}}+\varepsilon _{p\eta
}I_{33}+t_{p}(b-c)),
\end{equation}

\begin{eqnarray}
I_{\kappa_{s}p_{s}}=&& 6(-b-f+ d_{s} \alpha({\bf k})) -4b_{s}  \nonumber \\
&& - 3\left(c \frac{I_{\kappa_{s}\xi}} {I_{22}}+b \frac{I_{\kappa_{s}\eta}} {%
I_{33}}\right).
\end{eqnarray}

The parameters $b_{s}$, $f$, and $d_{s}$ are defined by

\begin{equation}
b_{s}=\left\langle d^{\alpha }(i)p_{s}^{\dagger }(i)\right\rangle ,
\end{equation}

\begin{equation}
f=\left\langle p^{\gamma}(i)d^{\dagger}(i)p^{\gamma }(i)p^{\gamma
^{\dagger}}(i)\right\rangle,
\end{equation}

\begin{equation}
d_{s}=\frac{1}{3}\left\langle n_{k}^{\alpha }(i)\sigma _{k}p^{\gamma
}(i)d^{\dagger }(i)\right\rangle .
\end{equation}

\bigskip


\end{multicols}

\begin{thebibliography}{99}

\bibitem[*]{byline}  Institute of Materials Science and Applied Research,
Sauletekio al. 9 - III, 2040 Vilnius, Lithuania

\bibitem{review1}  The Theory of Superconductivity in the High-Tc Cuprates
by P. W. Anderson, Princeton Series in Physics (1997).

\bibitem{review2}  High-Temperature Superconductivity: Experiment and Theory
by N.M. Plakida, Berlin Springer (1995).

\bibitem{anderson}  P. W. Anderson, Science, {\bf 235}, 1196 (1987); \hskip %
6pt Anderson, P. W., G. Baskaran, Z. Zou, and T. Hsu, Phys.\ Rev.\ Lett.\
{\bf 58}, 2790 (1987).

\bibitem{zhangrice}  F.C. Zhang and T.M. Rice, Phys.\ Rev.\ B {\bf 38}, 3759
(1988).

\bibitem{emery}  V. Emery and G. Reiter, Phys.\ Rev.\ B {\bf 38}, 4547
(1988).

\bibitem{barabanov1}  A.F. Barabanov, V.M. Beresovsky, L.A. Maksimov and E.
Zasinas, JETP, {\bf 83}, 819 (1996).

\bibitem{barabanov2}  A.F. Barabanov, L.A. Maksimov, O.V. Urazaev and E.
Zasinas, JETP letters, {\bf 66}, 182, (1997).

\bibitem{tobin}  J.G. Tobin {\it et al}, Phys.\ Rev.\ B {\bf 45}, 5563
(1992); \hskip 6pt R. Manzke {\it et al}, Surface Sci.\ {\bf 269}, 1066
(1992).

\bibitem{abrikos}  A.A. Abrikosov, J.C. Campuzano and K. Gofron, Physica C,
{\bf 214}, 73 (1993).

\bibitem{gofron}  K. Gofron {\it et al}, Phys.\ Rev.\ Lett.\ {\bf 73}, 3302
(1994).

\bibitem{wels}  B.O. Wells {\it et al}, Phys.\ Rev.\ Lett.\ {\bf 74}, 964
(1995).

\bibitem{mancini2}  S. Ishihara, H. Matsumoto, S. Odashima, M. Tachiki and
F. Mancini, Phys.\ Rev.\ B {\bf 49}, 1350 (1994).

\bibitem{mancini1}  F. Mancini, S. Marra and H. Matsumoto, Physica C {\bf 244%
}, 49 (1995); {\bf 250}, 184 (1995); {\bf 252}, 361 (1995).

\bibitem{mancini5}  F. Mancini, H. Matsumoto and D. Villani, Phys.\ Rev.\ B
{\bf 57}, 6145 (1998).

\bibitem{mancini6}  A. Avella, F. Mancini, D. Villani, L. Siurakshina, V.
Yu. Yushankhai, Int.\ Journ.\ Mod.\ Phys.\ B {\bf 12}, 81 (1998).

\bibitem{Adolfo}  F. Mancini and A. Avella, Condensed Matter Physics, {\bf 1}%
, 11 (1998).

\bibitem{becker}  K.W. Becker, W. Brenig and P. Fulde, Z.\ Phys.\ B {\bf 81}%
, 165 (1990); H. Mori, Progr.\ Theor.\ Phys.\ {\bf 33}, 423 (1965); ibid
{\bf 34}, 399 (1965); A.J. Fedro, Yu Zhou, T.C. Leung, B.N. Harmon and S.K.
Sinha, Phys.\ Rev.\ B {\bf 46}, 14785 (1992); P. Fulde, Electron
Correlations in Molecules and Solids (Springer-Verlag, Berlin-Heidelberg,
1993); N.M. Plakida, V.Yu. Yushankhai and I.V. Stasyuk, Physica C {\bf %
162-164}, 787 (1989); B. Mehlig, H. Eskes, R. Hayn and M.B.J. Meinders,
Phys.\ Rev.\ B {\bf 52}, 2463 (1995).

\bibitem{rowe}  D.J. Rowe, Rev.\ Mod.\ Phys.\ {\bf 40}, 153 (1968); L.M.
Roth, Phys.\ Rev.\ {\bf 184}, 451 (1969); J. Beenen and D.M. Edwards, Phys.\
Rev.\ B {\bf 52}, 13636 (1995).

\bibitem{kalashnikov}  O.K. Kalashnikov and E.S. Fradkin, Phys.\ Stat.\
Sol.\ (b) {\bf 59}, 9 (1973); W. Nolting, Z.\ Phys.\ B {\bf 255}, 25 (1972);
G. Geipel and W. Nolting, Phys.\ Rev.\ B {\bf 38}, 2608 (1988); W. Nolting
and W. Borgel, Phys.\ Rev.\ B {\bf 39}, 6962 (1989); A. Lonke, J.\ Math.\
Phys.\ {\bf 12}, 2422 (1971).

\bibitem{hmfm}  H. Matsumoto and F. Mancini, Phys.\ Rev.\ B {\bf 55}, 2095
(1997).

\bibitem{msm}  H. Matsumoto, T. Saikawa and F. Mancini, Phys.\ Rev.\ B {\bf %
54}, 14445 (1996).

\bibitem{mmvm}  F. Mancini, S. Marra, D. Villani and H. Matsumoto, Physics
Letters A {\bf 210}, 429 (1996).

\bibitem{mmv}  F. Mancini, S. Marra, and D. Villani, Condensed Matter
Physics {\bf 7}, 133 (1996).

\bibitem{ManciniBarabanov}  V. Fiorentino, F. Mancini, A.F. Barabanov,
Physica B {\bf 284}, 1195 (2000).

\bibitem{dagotto}  For a review see E. Dagotto, Rev.\ of Mod.\ Phys.\ {\bf 66%
}, 763 (1994).

\bibitem{Matsumotoxxx}  H. Matsumoto and M. Tachiki, Progr.\ Theor.\ Phys.\
Supplement {\bf 101}, 353 (1990).

\bibitem{mancini?}  F. Mancini and A. Avella, cond-mat/0006377.

\bibitem{McMahan}  E. B. Stechel and D. R. Jennison, Phys.\ Rev.\ B, {\bf 38}%
, 4632 (1988);\ A. K. McMahan, R. M. Martin and S. Satpathy, Phys.\ Rev.\ B,
{\bf 38}, 6650 (1989);\ M. S. Hybertsen, M. Schluter and N. E. Christensen,
Phys.\ Rev.\ B, {\bf 39}, 9028 (1989);\ J. B. Grant and A. K. McMahan,
Phys.\ Rev.\ B, {\bf 46}, 8440 (1992).

\bibitem{muramatsu1}  G. Dopf, A. Muramatsu and W. Hanke, Phys.\ Rev.\ B
{\bf 41}, 9264 (1990).

\bibitem{Scalettar}  R.T. Scalettar, D.J. Scalapino, R.L. Sugar and S.R.
White, Phys.\ Rev.\ B {\bf 44}, 770 (1991).

\bibitem{muramatsu2}  G. Dopf, J. Wagner, P. Dieterich, A. Muramatsu and W.
Hanke, Phys.\ Rev.\ Lett.\ {\bf 68}, 2082 (1992).

\bibitem{Birgeneau}  R.J. Birgeneau {\it {et al.}}, Phys.\ Rev.\ B {\bf 38},
6614 (1988).

\bibitem{mancinisusceptibility}  F. Mancini, H. Matsumoto and D. Villani,
Journ. of Physical Studies {\bf 3}, 474 (2000).

\bibitem{ino}  A. Ino, C. Kim, T. Mizokawa, Z.-X. Shen, A. Fujimori, M.
Takaba, K. Tamasaku, H. Eisaki, S. Uchida, J. Phys. Soc. Jpn. {\bf 68}, 1496
(1999).

\bibitem{hall}  T. Plackowski and M. Matusiak, Phys.\ Rev.\ B {\bf 60},
14872 (1999).

\end{thebibliography}
\end{document}